\documentclass[useAMS,usenatbib]{mn2e}
\usepackage{graphics,graphicx}
\newcommand{\kms}{km s$^{-1}$}

\def\nat{Nature}

\def\apj{ApJ}

\def\apjs{ApJS}
\def\aap{A\&A}

\def\mnras{MNRAS}
\def\araa{ARA\&A}
\def\qjras{QJRAS}

\title[Dense gas in AFGL 437]{Dense gas and exciting sources of the
  molecular outflow  in the 
AFGL 437 star-forming region}
\author[G. Manjarrez et al.]{G. Manjarrez$^{1,2,3}$\thanks{E-mail:
gmanjarr@eso.org}, J. F. G\'omez$^{3}$\thanks{On sabbatical leave at  CSIRO Astronomy \& Space Science, Marsfield, NSW 2122, Australia} and
I. de Gregorio-Monsalvo$^{1,4}$\\
$^{1}$European Southern Observatory, Alonso de C\'ordova 3107, Vitacura, Casilla 19001, Santiago 19, Chile\\
$^{2}$Centro de Radioastronom\'ia y Astrof\'isica, UNAM, Morelia, 58089, Mexico\\
$^{3}$Instituto de Astrof\'{\i}sica de Andaluc\'{\i}a, CSIC, Apartado
3004, E-18080 Granada, Spain\\
$^{4}$ Joint ALMA Observatory, Alonso de C\'ordova 3107, Vitacura, Santiago, Chile \\
  }
\begin{document}

\voffset=-0.8in

%\date{Accepted xxx. Received xxx; in original form xxx}

%\pagerange{\pageref{firstpage}--\pageref{lastpage}} \pubyear{2002}

\maketitle

\label{firstpage}

\begin{abstract}

We  present  Very Large Array (VLA) high resolution observations of the NH$_3$(1,1) 
and NH$_3$(2,2) molecular transitions towards the high mass star 
forming region AFGL 437. Our aim was to investigate if the 
poorly collimated CO molecular outflow  previously detected in the 
region is the result of a projection effect, with no intrinsic 
bipolarity, as suggested by G\'omez et al. We complemented 
our observations with radio continuum archived data from the VLA
 at 2 and 3.6 cm, and with unpublished public data at 450  $\mu$m
 taken with Submillimetre Common-User Bolometer Array at the James Clerk Maxwell Telescope. 
Ammonia emission was 
found  mainly in three clumps located at the  south and east of 
the position of the compact infrared cluster of AFGL 437, where the CO outflow seemed to have its origin. One of the   NH$_3$(1,1) clumps  coincides with  the 
maximum  of NH$_3$(2,2) and  with a local peak of emission at 450 $\mu$m.
 A near infrared source (s11) is also found at that position.  Our 
continuum map at 2 cm shows  extended elongated emission associated 
with the infrared source AFGL 437W. This elongated morphology and 
its spectral index  between 3.6 and 2 cm ($\simeq 0.4$) suggest the presence of a jet in 
AFGL 437W. We suggest that  several molecular bipolar outflows may 
exist in the region. The observed CO outflow would be the superposition 
of those individual outflows, which would explain its low degree of 
collimation observed at larger scales.

\end{abstract}


\section{Introduction}

Mass-loss phenomena are one of the best-known manifestations of the
star formation process. In our current understanding of the formation of
low-mass stars via accretion (e.g., McKee \& Ostriker 2007), 
highly collimated jets are necessarily present in the first
stages of protostellar evolution, coeval with the formation of
circumstellar disks. Jets could be the agent releasing angular momentum
excess, so that the protostar can continue accreting material from its
environment.

Evidence for highly collimated mass-loss is widespread in low-mass young
stellar objects (YSOs): e.g., jets
traced by Herbig-Haro objects, radio continuum emission, or masers, as
well as bipolar molecular outflows. 
However the case for high-mass stars ($M\geq 8$ M$_{\odot}$) 
is less clear. Energetic
mass loss is indeed present in those sources, although it is, in
general, less collimated than in low-mass objects \citep{Wu-etal04}. 
Highly collimated jets seems to be
restricted to the earliest phases of the evolution ($<10^4$ years;
Shepherd 2005) of massive YSOs.

This is important from a more global perspective, since it is not yet
clear whether high-mass stars form via accretion, like their low-mass
counterparts (e.g., Yorke \& Sonnhalter 2002; McKee
\& Tan 2003; Krumholz et al. 2009), or by coalescence of
lower-mass objects (Bonnell, Bate \& Zinnecker 1998). 
The presence of circumstellar disks and collimated outflows are key
ingredients of the accretion scenario. The detection of disks 
\citep{Patel-etal05} and jets  (e.g., Mart\'{\i}, Rodr\'iguez \& Reipurth 1993; 
Rodr\'{\i}guez et al. 1994; Davis et al. 2004; Patel et al. 2005) in several
high-mass objects clearly indicate that formation via accretion is
possible in these objects, but we still do not know if this process is of
general application.

\begin{figure*}
 \begin{center}
 \rotatebox{0}{
\includegraphics[scale=0.7]{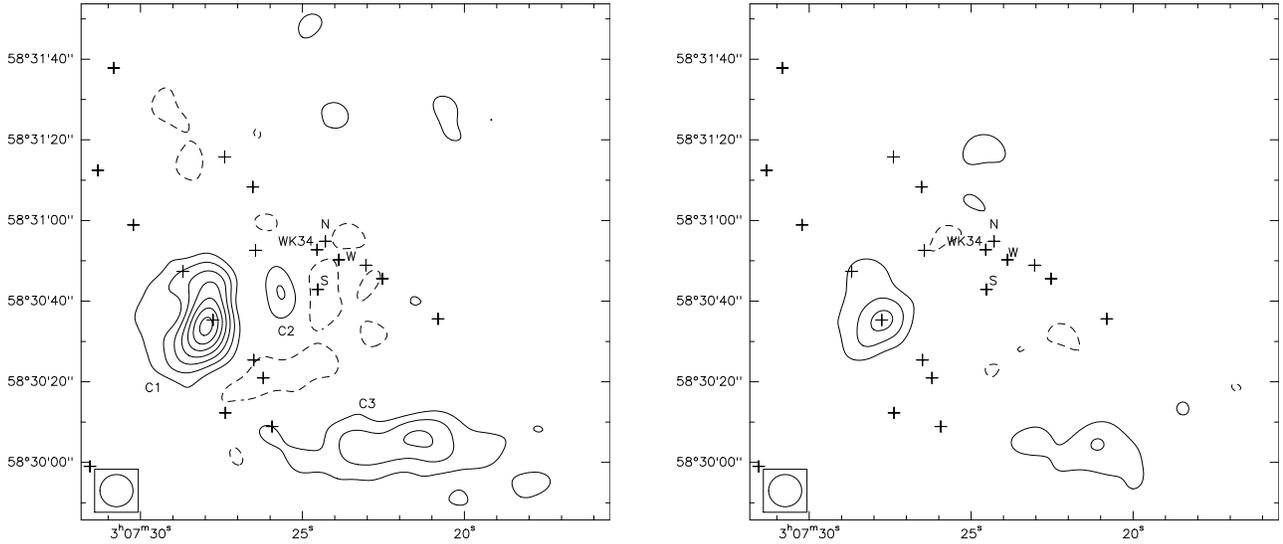}}
 \caption{\label{test1} Integrated intensity map of  the main hyperfine component of the NH$_3$(1,1)  (left) and NH$_3$(2,2) (right) 
lines. The lowest contour
   levels and the increment step are 3 times 8.7 mJy  km s$^{-1}$, the rms of the
   map (beam$=8''$,  shown at the bottom left-hand corner of each map). Ammonia 
   clumps are labelled as C1, C2, and C3. Crosses mark the
   position of the infrared sources detected with Spitzer \citep{Dewangan&Anandarao10},
 of which the better studied ones are labelled.  Axes are right
 ascension and declination in equinox J2000.}   
 \end{center}
\end{figure*}

Here we will study the AFGL 437 region, which hosts an interesting
molecular outflow. The AFGL 437 region comprises a cluster of at 
least $\sim 20$ YSOs \citep{Weintraub&Kastner96,Dewangan&Anandarao10}, 
although infrared images are dominated by a compact ($\sim 15''$) central cluster of
four sources (named AFGL 437N, S, E, and W by Wynn-Williams et al. 1981), which seem to  
have recently emerged from the near side of the molecular cloud.  
AFGL 437 shows clear signs of ongoing star formation, such as the 
presence of radio continuum emission (probably tracing ultracompact HII
regions associated with sources W and S, Wynn-Williams et al. 1981;
Torrelles et al. 1992; Kurtz, Churchwell \& Wood 1994), water masers (close to 
sources N and W, Torrelles et al. 1992), and a molecular outflow traced
by CO \citep{Gomez-etal92}. At least sources W and S are thought to
be massive stars, of early B type \citep{WynnWilliams-etal81,
Torrelles-etal92}.

The molecular outflow, observed with single dish at $13''$ angular
resolution  \citep{Gomez-etal92}, is roughly oriented in the north-south
direction, but it shows a very low degree of
collimation, with high-velocity CO emission completely surrounding the
central cluster. Interestingly, all observed CO isotopes show the same
distribution of blue- and red-shifted gas. Given the location of the
YSOs close to the edge of the parental cloud, \cite{Gomez-etal92} suggested a
possible interpretation for the nature of the molecular outflow,
alternative to more classical models: if the winds from the YSOs, that
can be isotropic at the origin, shock obliquely against the walls of
the cavity opened in the cloud, they can produce a laminar flow of
molecular gas along these walls. When this flow is observed under a
particular angle of view, it may give the appearance of a bipolar
molecular outflow. The observed bipolarity would then be a
projection effect, not an intrinsic characteristic of the mass-loss
itself. An alternative explanation for the morphology of the molecular
outflow would be the superposition of several outflows excited by
different sources in the cluster.

The IR observations showed two different sources 
towards AFGL 437N \citep{Rayner&McLean87,Weintraub&Kastner96}, 
one of which (WK 34) is associated with a bipolar polarized nebula 
(Weintraub \& Kastner 1996; Meakin, Hines \& Thompson 2005) oriented N-S (i.e., roughly 
in the same direction as the CO outflow) with a centrosymmetric polarisation 
pattern. This led  \cite{Weintraub&Kastner96} to suggest that
WK 34 is the main driving source of the molecular outflow. However,
the nature of this source is still uncertain: while \cite{Meakin-etal05}
 argued that it is a low-mass YSO, recent
SED modelling including Spitzer data \citep{Dewangan&Anandarao10}
suggest that it could be massive, but of young age, with effective
temperature still not sufficient to create an HII region. 
On the other hand, high-resolution infrared observations show elongated
(monopolar) emission associated with source AFGL 437S \citep{Alvarez-etal04}.
 This suggests that more than one source in the region 
could be undergoing mass loss.

In this paper we present high resolution ammonia observations of the AFGL 437
region, designed to test whether the bipolarity of the outflow is 
as a projection
effect (as suggested by G\'omez et al. 1992) 
or if, alternatively, the observed outflow is the superposition
of individual outflows from different sources. 
In the case of a projection effect, we should expect the ammonia to
show a spatio-velocity pattern similar to that of CO isotopes, with
blueshifted gas to the south and redshifted one to the north. On the
other hand, if this is a more classical example of molecular outflow,
we expect ammonia emission to peak near the exciting source of the
outflow, and possibly tracing a disk/toroid 
elongated in the E-W direction, i.e.,
perpendicular to the collimation axis (e.g., Estalella et al. 1993;
Wiseman et al. 2001).

We complement the
ammonia observations with
archival data of continuum emission at 
centimetre and submillimetre wavelengths to further study the nature
of the sources
in this region.

\begin{table*}
 \centering
 \begin{scriptsize}
  %\begin{minipage}{\textwidth}
    \caption{Observed properties of NH$_3$(1,1) clumps.\label{obsparams}}
    \begin{tabular}{p{0.5cm}p{1.2cm}p{1.2cm}p{1.5cm}p{1.5cm}p{1.5cm}p{1.5cm}p{1.4cm}p{1.5cm}p{1.2cm}}
      \hline
      Clump & R.A.$^1$ & Dec $^1$& $I(1,1;m)$ $^2$ &
      $I(1,1;s)$ $^3$  &
      $V_1$ $^4$ & $\Delta V_1$ $^5$ & $I(2,2;m)$ $^6$ & $V_2$ $^7$ &
      $\Delta V_2$ $^8$ \\
            &     $_\mathrm{(J2000)}$    &  $_{\mathrm{(J2000)}}$       & $_{\mathrm{(mJy~beam}^{-1})}$      &  $_{\mathrm{(mJy~beam}^{-1})}$ &
       $_\mathrm{(km~s^{-1})}$   & $_\mathrm{(km~s^{-1})}$  & $_{\mathrm{(mJy~beam}^{-1})}$  & $_\mathrm{(km~s^{-1})}$ & $_\mathrm{(km~s^{-1})}$  \\
      \hline
      C1 & 03 07 27.9 & 58 30 34 & $53.3\pm 1.1$ & $18.3\pm 1.1$ &
      $-41.11\pm 0.02$ & $2.21\pm 0.05$ & $23.4\pm 0.6$ &
      $-41.04\pm 0.03$ & $1.96\pm 0.07$\\
      C2 & 03 07 25.7 & 58 30 42 & $28.8\pm 1.0$ & $12.2\pm 1.1$ &
      $-40.00\pm 0.02$ & $1.29\pm 0.05$ & $17.6\pm 1.0$ &
      $-40.19\pm 0.03$ & $0.85\pm 0.06$\\
      C3 & 03 07 21.4 & 58 30 06 & $36.7\pm 1.1$ & $13.1\pm 1.7$ &
      $-40.43\pm 0.02$ & $1.34\pm 0.05$ & $18.7\pm 1.1$ &
      $-40.42\pm 0.03$ & $1.05\pm 0.07$\\
      \hline
    \end{tabular}
\begin{minipage}{\textwidth}
 \medskip
    1. Coordinates of the positions of the local peak in the
    integrated intensity map of NH$_3$(1,1) (see Fig. \ref{test1}). Units
    of right ascension are hours, minutes, and seconds; units of
    declination are degrees, arcminutes, and arcseconds.\\
    2. Intensity of the  main hyperfine component of the NH$_3$(1,1) line, at
    that position. \\
    3. Intensity of the inner satellite component of  NH$_3$(1,1).\\
    4. LSR velocity of the NH$_3$(1,1) line.\\
    5. Full width at half maximum (FWHM) of the  main hyperfine component of the NH$_3$(1,1) line. 
     It was derived from a simple Gaussian fit, without considering the magnetic hyperfine 
    structure   within the lines. \\
    6. Intensity of the main hyperfine component of the NH$_3$(2,2) line.\\
    7. LSR velocity of the NH$_3$(2,2) line.\\
    8. FWHM of the main hyperfine component of the NH$_3$(2,2) line. 	
  \end{minipage}
\end{scriptsize}
\end{table*}

\section{Observations}

We observed the (1,1) and (2,2) inversion transitions of the NH$_3$
molecule (rest frequencies 23694.496 and 23722.634 MHz, respectively)
towards the AFGL 437 region, using the Very Large Array (VLA) of the National Radio Astronomy 
Observatory\footnote{The National Radio Astronomy Observatory is a facility of
the National Science Foundation operated under cooperative agreement by
 Associated Universities, Inc.} in its D
configuration on 2004 August 16 (project AG665). The phase centre was 
located at R.A.(J2000) = $03^h 07^m 23.7^s$, Dec(J2000) = $58^\circ
30' 50''$. We observed both lines simultaneously in dual circular polarisation, 
using the 4IF mode of the VLA. This allowed us to sample each line
with 63 spectral channels of 48.8 kHz (0.62 km s$^{-1}$) width,
centred at $V_{\rm LSR}$ = $-39.4$ km s$^{-1}$. The flux calibrator
was J0137+331, with an assumed flux density of 1.05 Jy using the latest VLA values (1999.2). 
The phase and bandpass calibrator was J0359+509, with bootstrapped flux density of
 9.5 Jy. We calibrated and imaged the data using
the standard procedures of the Astronomical Image Processing System
(AIPS) of the National Radio Astronomy Observatory (NRAO). Continuum
emission was subtracted using task UVLIN. Mapping and deconvolution
was carried out with task IMAGR, using a natural weighting of the
visibilities. To improve the signal-to-noise ratio for extended
emission we convolved the resulting images with a Gaussian beam, to
obtain a final synthesised beam size of $8''$.

We have also reprocessed archival radio continuum data of this region
at 2 and 3.6 cm. These observations correspond to VLA projects AC240
(1989 March 19, B configuration), AT121
(1991 May 25, D configuration), and 
AT122 (1991 January 30, CnD configuration), and were published in
Torrelles et al. (1992; projects 
AT121 and AT122) and 
Kurtz et al. (1994; project AC240). The description of those observations can be
found in these papers. 
Further self-calibration was possible for the data observed in
1991. We combined
together the visibility data at each wavelength before imaging. 
With this combination we can
properly sample the emission at both, larger  (as traced by the
Torrelles et al. data), and smaller scale (given by the
higher-resolution Kurtz et al. data). Imaging of the combined data was
carried out with a robust weighting of the visibilities, with robust
parameters 0 and 5 for the data at 3.6 and 2 cm, respectively. With
these parameters, the final data have a similar size of their
synthesised beams ($1.01''\times
0.84''$ at 3.6 cm, and $1.18''\times 1.11''$ at 2 cm). 

\begin{figure}
 \begin{center}
 \rotatebox{-90}{\includegraphics[height=84mm]{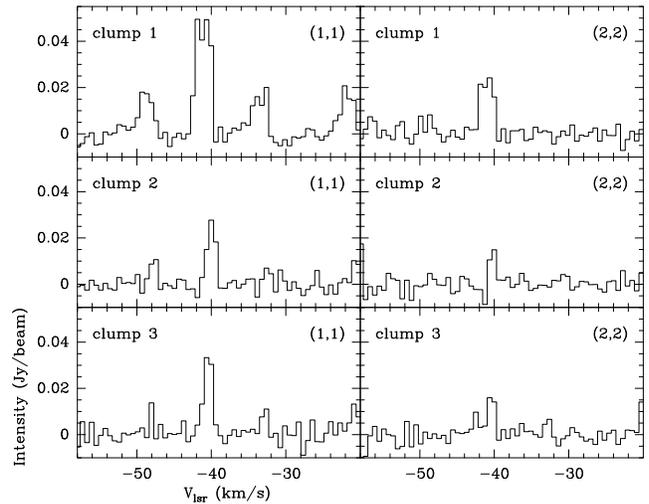}}
 \caption{\label{spectra} Spectra of the NH$_3$(1,1) and (2,2) lines
   at the position of the peaks of the clumps shown in Fig. \ref{test1}.}   
 \end{center}
\end{figure}

We have also retrieved publicly available submillimetre data at 450
and 850 $\mu$m, taken with SCUBA at the James Clerk Maxwell Telescope
 (JCMT)\footnote{The James Clerk Maxwell Telescope is operated by the 
Joint Astronomy Centre on behalf of the Science and Technology Facilities 
Council of the United Kingdom, the Netherlands Organisation for
Scientific Research, and the National Research Council of Canada.} on 2003
November 6, corresponding to project M03BU13. The data at 850 $\mu$m
have been presented in \cite{Curran&Chrysostomou07} and \cite{Matthews-etal09}, 
but in this paper we are using for our analysis the original data 
($\simeq 14''$ angular resolution), without applying any spatial
convolution (final angular resolution of $20''$ in Matthews et
al. 2009). The data at 450 $\mu$m are presented here for first time. The 
angular resolution of the JCMT at 450 $\mu$m is $\simeq 8''$.

\begin{figure*}
 \begin{center}
 \rotatebox{0}{
 \includegraphics[scale=0.70]{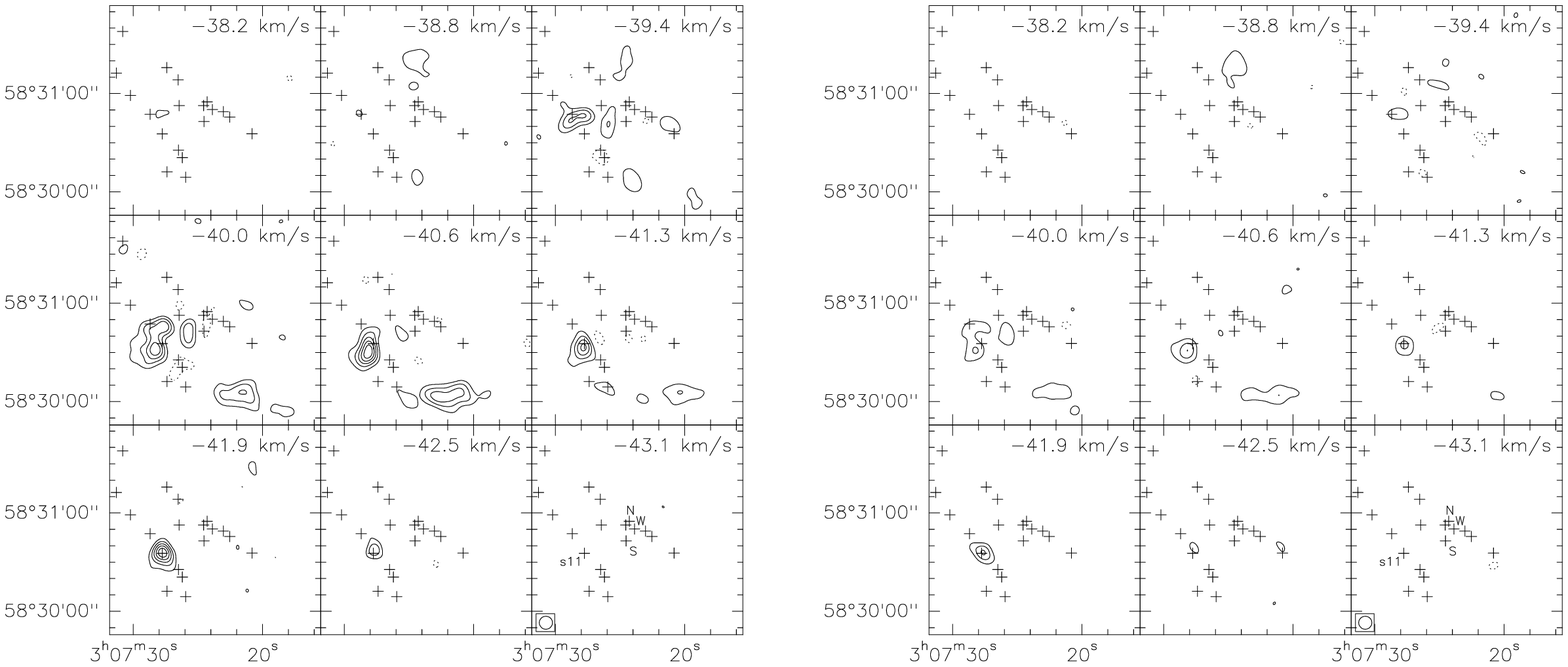}}
 \caption{\label{chan} Channel maps of the  main hyperfine component
   of the NH$_3$(1,1)  (left) and NH$_3$(2,2) (right) lines. The lowest
 contour level and the increment  step are 3 times 3.3 mJy, the rms of the maps
 (synthesized beam$=8''$,  shown at the bottom left-hand corner of the
 channel with velocity $-43.1$ \kms).}   
 \end{center}
\end{figure*}

\section{Results}  

\subsection{Ammonia observations}

Fig. \ref{test1} shows the
integrated intensity maps of the  main hyperfine component of NH$_3$(1,1) and (2,2) lines.
 No significant
emission is present at the central cluster of stars. Two 
clumps are clearly present in both maps, to the east and south of the
centre (labelled as C1 and C3 in the figures). 
Their morphology appears elongated, specially in the (1,1) map.
The spectra
toward the peaks of those clumps are shown in Fig. \ref{spectra}.  The main and inner 
satellite components of the electric quadrupole hyperfine structure of the (1,1) transition 
are evident.
A weaker clump (labelled C2) is also visible in the
(1,1) map, west from C1. In
tables \ref{obsparams} and \ref{phyparams} we list the observed and the derived physical
parameters, respectively, of those
three clumps.  Line intensities, central positions and widths in Table
\ref{obsparams} were obtained     with Gaussian fits to the emission.
 Physical parameters in Table \ref{phyparams} were obtained following the 
formulation of \cite{Ho&Townes83}, and \cite{Mangum-etal92}.

Clump C1 (the most intense one in NH$_3$) appears to have the highest
column density. Moreover, its lines are wider than the rest. 
There is an infrared source, detected in Spitzer IRAC bands (s11,
Kumar Dewangan \& Anandarao 2010), close to 
its maximum, at the exact position of the NH$_3$(2,2) peak. 
These characteristics may imply that C1  harbours a protostar inside.
No infrared source appears clearly associated to the peak
of clumps C2 and C3.

Channel maps are shown in Fig. \ref{chan} for the
(1,1) and (2,2) lines. We do not see any obvious velocity pattern within the
region, although there is a
hint of redshifted emission to the north (see panel at $-38.8$ km
s$^{-1}$). This would suggest a trend (redshifted to the north,
and blueshifted to the south) similar to that found in the CO outflow
\citep{Gomez-etal92}, although the low signal to noise ratio of the emission,
apart from clumps C1 and C2 precludes any firm
conclusion based on the velocity pattern.

\begin{table}
    \caption{Physical parameters of NH$_3$(1,1) clumps.\label{phyparams}}
    \begin{tabular}{llllll}
      \hline
      Clump & $\tau_{11}$ $^1$ & $T_R$ $^2$& $N_{H_2}$ $^3$ & size
      $^4$& $M$ $^5$\\
            &             & (K)  & ($10^{22}$ cm$^{-2}$) & (arcsec) & (M$_\odot$)\\
      \hline 
      C1 &  0.6 & 18 & $1.4$ & $21\times 12$ & 5\\
      C2 &  1.4 & 20 & $0.5$ &  $11\times 8$    & 0.7 \\
      C3 & 0.8  & 19 & $0.5$ & $40\times 13$   & 4\\
      \hline
    \end{tabular}

    \medskip
    1. Optical depth of the  main hyperfine component of the
    NH$_3$(1,1) line.\\
    2. Rotational temperature.\\
    3. Column density of hydrogen, assuming local thermodynamic equilibrium, with kinetic and excitation temperature of NH$_3$ lines equal to $T_R$, and  an abundance of
    NH$_3$ relative to hydrogen of $10^{-8}$.\\
    4. FWHM size of the clump.\\
    5. Estimated mass of the clump.
 \end{table}

We note that single-dish observations (obtained with the Effelsberg antenna, 
angular resolution $\simeq$
40''; Wu et al. 2006) show ammonia emission associated with the central
cluster, specially in the (2,2) transition. Since we do not detect
this in our observations, it is possible that the emission detected by
\cite{Wu-etal06} is extended and relatively uniform, so that either it
is below the sensitivity limit when observed at
$8''$ resolution, or a fraction of it is missed because of the lack of short
interferometer spacings.  To estimate the amount of emission that 
our interferometric observations are missing, we have convolved our integrated intensity maps with an elliptical Gaussian, to obtain a final angular 
resolution of $42.2''$, corresponding to that of Effelsberg at the ammonia frequency. We then compared the integrated intensities tabulated by Wu et al. 
(2006) with the values in our convolved map. At the position given by Wu 
et al. (2006), the VLA is recovering $\simeq 12$\% and $33$\% of the NH$_3$(1,1) and (2,2) emission, respectively. Therefore, we are missing a significant 
fraction of the NH$_3$ emission, which would complicate any conclusion based on large-scale structures. However, at smaller scales, 
we are confident that the individual ammonia 
clumps seen in 
Fig. 1 are real structures, given their good correspondence with 
submillimeter emission (see next section). To get a complete picture of 
the overall distribution of ammonia map, we should combine both 
interferometric and single-dish data. Unfortunately, the available 
single-dish data are not enough to properly carry out such a combination, 
since the Wu et al. (2006) maps are not fully sampled (their observed 
points have full-beam separation).

\subsection{Submillimeter continuum emission}

\begin{figure*}
 \begin{center}
 \rotatebox{-90}{
 \includegraphics[width=84mm]{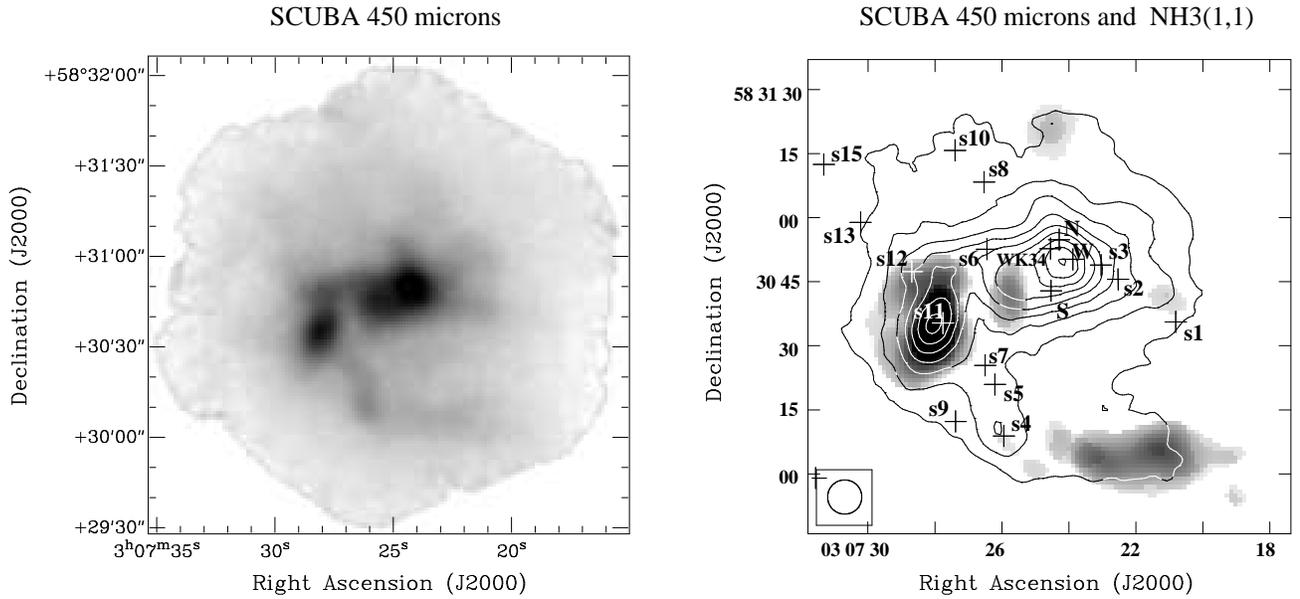}}
 \caption{\label{scuba}(Left)  Map of the emission at 450 $\mu$m. (Right) An 
overlaid of 450 $\mu$m emission (contours) with integrated emission
 (greyscale) of the main component of the NH$_3$(1,1)  line.  Contours are -5, 5, 10, 15, 20, 25, 30, 35, 40, 45 and 50 times 0.152 Jy beam$^{-1}$, the rms of the map (beam$\simeq8''$,  shown at the bottom left-hand corner of the right map).}   
 \end{center}
\end{figure*}

The SCUBA maps at 850 $\mu$m presented in \cite{Curran&Chrysostomou07} and \cite{Matthews-etal09} showed extended, elongated
emission oriented in the northwest-southeast direction. However, in the maps at 450 $\mu$m,
(Fig. \ref{scuba}), we can clearly
distinguish two local maxima. One of them is located at the position
of the central cluster, and the other one coincides with C1 (with its
maximum located at R.A.(J2000) $= 03^h07^m28.1^s$, Dec(J2000)
$=58^\circ 30' 35''$). We
also note that the central submillimeter emission extends to the east,
toward C2. There is also a weaker finger of emission elongated
toward the south that peaks at the position of the IR source s4 (see Fig. \ref{scuba}), and extended, very faint emission coincident with C3.
 Apart from the central peak, the submillimeter emission tends to
follow the distribution of  NH$_3$. This gives
us confidence that the ammonia {clumps  seen in Fig. \ref{test1} are real entities, and not artifacts generated by the lack of short spacings in the interferometer}. 

From the map at 450 $\mu$m, we can estimate the mass of the observed submillimeter
clumps, using the formula 
$M=S_\nu d^2/(\kappa_\nu B_\nu (T_d))$,
where $S_\nu$ is the flux density, $d$ is the distance
to the source (2 kpc), $\kappa_\nu$ is the mass absorption coefficient,
and $B_\nu(T_d)$ is the Planck function for the dust temperature,
$T_d$. The value of $\kappa_\nu$ is highly uncertain, and is very
sensitive to the particular dust properties of the object. We have
adopted the prescription suggested by \cite{Hildebrand83}, $\kappa_\nu =
0.1 (250 \mu{\rm m}/\lambda)^\beta$ cm$^2$ g$^{-1}$, which assumes a
gas to dust ratio of 100. To estimate the dust emissivity index
($\beta$) we convolved the 450 $\mu$m map with an elliptical gaussian,
to match the angular resolution of the one at 850 $\mu$m. After this
convolution, the ratio between those maps is $\simeq 9$, which yields
$\beta\simeq 2.2$, a value within the range found in other high-mass
stars \citep{Molinari-etal00}. With these parameters, flux densities of
6.9 and 5.6 Jy represent masses of 44 and 39 M$_\odot$ for the central
submillimeter peak and the one associated with the ammonia clump C1,
respectively. In the latter case, the mass obtained from the
submillimeter data is larger than the one from ammonia (5 M$_\odot$),
but consistent with our estimate that the ammonia data may be
recovering only $\simeq 12$\% of the total flux.

\subsection{Centimeter continuum emission}

\begin{figure*}
 \begin{center}
 \rotatebox{-90}{
 \includegraphics[scale=0.65]{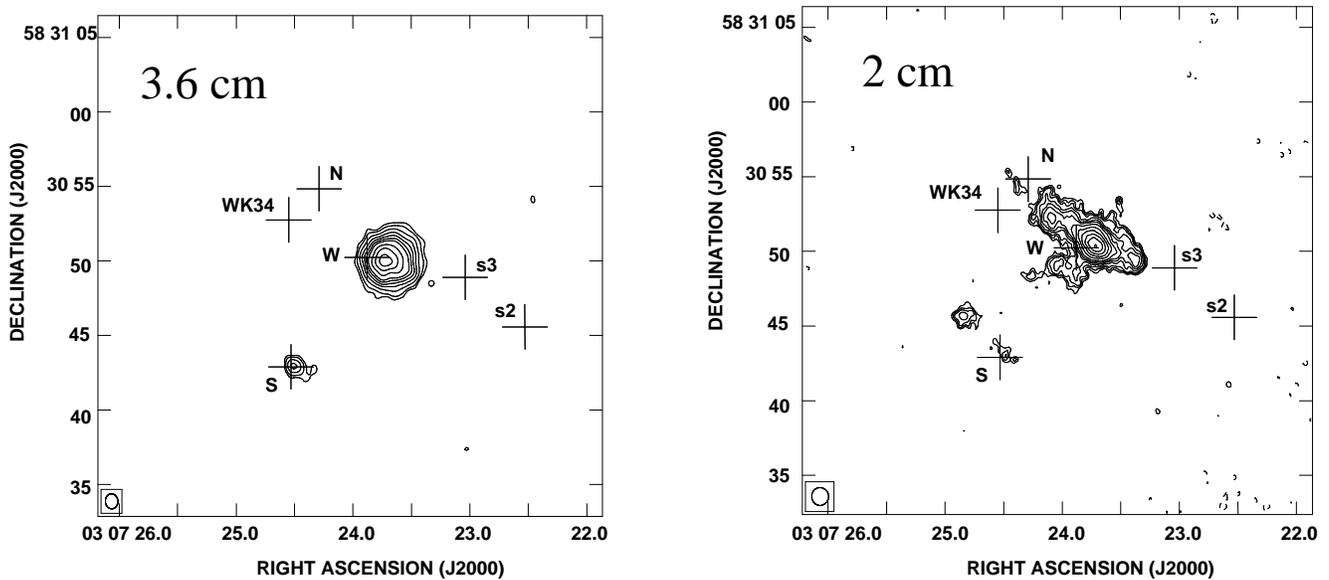}}
 \caption{\label{radio3} (left) Radiocontinuum emission at 3.6 cm in the 
central cluster of AFGL 437. Contour levels are -4, 4, 6, 10, 15, 20, 25, 
30, 40, 50 and 60 times 0.042 mJy beam$^{-1}$, the rms of the map (beam$=1.01''\times0.84''$, shown at the bottom left-hand corner of the map). (Right) Radiocontinuum  
emission at 2 cm. Contours are -4, 4, 5, 6, 8, 10, 12, 15, 20, 25, 30, 35 
and 40 times 0.1034  mJy beam $^{-1}$, the rms of the map (synthesized beam$=1.18''\times1.11''$).}   
 \end{center}
\end{figure*}

Maps of the continuum at 3.6 and 2 cm are shown in Fig. \ref{radio3}.
 As reported previously \citep{WynnWilliams-etal81,Torrelles-etal92,Kurtz-etal94},  there
are two main radio continuum sources, as seen in Fig. \ref{radio3},
associated with sources AFGL 437W and S. Parameters of the continuum
sources are shown in Table \ref{tabcont}. However, our map at 2 cm
 shows extended, elongated emission in AFGL 437W,
in the NE-SW direction (P.A. $\simeq 60^\circ$). 
This elongated emission was
not  clearly seen in 
the individual datasets that we combined together to obtain this map.
 It is possible that the elongated structure only shows up with good
enough coverage of the uv-plane, 
which would also explain why it is not seen in the
3.6 cm data. If this interpretation is correct, 
further observations at 3.6 cm with a better uv-coverage
could detect the elongated structure.

\begin{table}
    \caption{Centimeter radio continuum sources.\label{tabcont}}
    \begin{tabular}{llll}
      \hline
      Source & $S(3.6cm)$ $^1$& $S(2cm)$ $^2$& $\alpha(3.6-2)$ $^3$\\
            &    (mJy)   & (mJy)  & \\
      \hline 
      W &  $22.1\pm 0.7$ & $28.3\pm 2.2$ & $0.4\pm 0.2$ \\
      S &  $1.5\pm 0.4$ & $<2$ & $<0.5$  \\
      \hline
    \end{tabular}

\medskip
1. Flux density at 3.6 cm.\\
2. Flux density at 2 cm.\\
3. Spectral index between 3.6 and 2 cm, defined as $S_\nu \propto \nu^\alpha$.
\end{table}

The elongated radio continuum emission suggests the presence of a jet,
and it is reminiscent of other jets traced by radio continuum in
high-mass star-forming regions, like Cepheus A or W75N (Torrelles et
al. 1996, 1997). The derived spectral index of this source
($0.4\pm 0.2$) is 
similar to that expected from constant-velocity winds ($\simeq 0.6$).
Following the formulation by \cite{Reynolds86}, the spectral index would
indicate the presence of a confined jet (i.e, narrower than a purely
biconical shape). This formulation assumes a power-law dependence of width ($w$) with distance from the
center ($r$), of the form $w\propto r^\epsilon$. In our case, the
derived index is  $\epsilon
\simeq 0.8$.

The mass loss rate of the ionized jet can be estimated from eq. 19 of
Reynolds (1986). We assumed a wind velocity of 500 km s$^{-1}$, 
a turnover frequency
for the  radio continuum emission of 10 GHz, a temperature of the ionized gas of $10^4$
K, a jet opening angle of $\simeq 30^\circ$, and an inclination angle
with respect to the line of sight of $45^\circ$.  We have chosen a
turnover frequency between the two observed ones, since the derived
spectral index (0.4) is partially optically thick. However, the value
of the mass loss rate depends only weakly on the turnover frequency
(to the power of $-0.15$), and the possible error introduced by this value is
small. Under these
assumptions, we obtain a mass-loss rate of $\simeq 5\times 10^{-6}$
M$_\odot$ yr$^{-1}$. The assumed velocity of 500 km s$^{-1}$ would
imply a momentum rate of $\simeq
2.5\times 10^{-3}$ M$_\odot$ km s$^{-1}$  yr$^{-1}$ for this jet, which
could be enough to 
drive the large scale CO outflow in the region,  given that the derived
momentum rate of jet and outflow are of the same order (G\'omez et al. 1992).

 We also note that
mid-infrared images of this  source \citep{deWit-etal09} show extended emission
slightly elongated along the same direction.

In the 2 cm map, there seems to be also some radio continuum emission $\simeq 3.7''$  
northeast of AFGL 437S, and an extension to the southeast of AFGL 437W. 
We have searched in data archive for possible infrared counterparts of 
those emissions, but none was found. At this point, we cannot ascertain 
whether these represent two real sources, or they are instrumental 
artifacts. If they are indeed real sources, the lack infrared counterpart 
for the weak 2 cm continuum emission may suggest the presence of young, 
deeply embedded sources. Some weak continuum emission may also be associated with AFGL 437N.

\section{Discussion}

G\'omez et al. (1992) suggested an interpretation for the molecular
outflow in terms of motions of gas along the wall of a cavity opened
in the molecular cloud by the massive stars in the cluster. With this
anisotropic distribution of molecular gas, even if the stellar winds
are isotropic, the outflow could appear as bipolar, and of low
collimation, to the observer.

Our NH$_3$ observations were specifically designed to test whether
this ``alternative'' interpretation for the outflow is valid, or this
is a more ``classical'' bipolar outflow, where there is intrinsic
collimation in the mass-loss process, probably driven by source WK
34. There are several key characteristics
to be investigated with these observations, such as the velocity pattern, the
morphology, and the location of NH$_3$.

\subsection{The velocity pattern of NH$_3$} 

If the interpretation
  given by \cite{Gomez-etal92} for the outflow, as motions along the
  walls of a cavity is correct, this drag of material should be seen
  in all molecular tracers. In particular, the spatial distribution of
  blue- and redshifted velocities of the dense gas traced by NH$_3$
  should be similar to the one seen in CO. 

Our results (Fig.\ \ref{chan}) are
  not conclusive to reject the G\'omez et al.'s hypothesis. There is
  some hint of more redshifted gas to the north, and more blueshifted
  one to the south, but the low signal-to-noise ratio of the emission,
  specially that to the north does not allow firm conclusions. On the
  other hand, the most intense ammonia emission (C1) seems to be related to
  a particular infrared source (s11), which also emits in the
  submillimeter. The ammonia seems to trace individual clumps, well
  defined in velocity, rather than a continuous distribution with a
  velocity gradient, which one could expect from the model proposed by
  \cite{Gomez-etal92}.

\subsection{The morphology  and location of NH$_3$} 

Classical examples
  of bipolar outflows show interstellar toroids of dense gas, perpendicular to
  molecular outflows and/or jets (e.g., Torrelles et al. 1983; Wiseman
  et al. 2001). In the past, these interstellar toroids have been
  proposed as the collimating agents of these outflows, although the
  actual collimators seem to be much smaller circumstellar
  disks. However, given the usual relationship outflow-interstellar
  toroid, the presence of a dense structure (traced with NH$_3$)
  perpendicular to the outflow, and located close to is center, 
  would have favored a more   ``classical'' interpretation of the outflow. 
 Moreover, ammonia emission has been used as a tool to identify the excitation
source of bipolar outflows. Excitation source of outflows usually
coincides with the maximum emission of ammonia 
\citep{Anglada-etal89}. This is usually confirmed with the presence
of local enhancements of temperature (with a higher ratio of the
emission of ammonia
(2,2) to (1,1) transitions), and turbulence (wider ammonia lines).

The overall distribution of  NH$_3$ does not show any preferential
orientation with respect to the molecular outflow. Only clump 3 shows
an E-W orientation, roughly perpendicular to the outflow. However, it
is located to the south, on the blueshifted lobe of the CO emission,
rather than towards the central cluster, and no infrared source is
located near its maximum. 

Using the maximum of ammonia (clump C1) as a criterium to search for the
powering source of the outflow, source s11 would be a good
candidate. 
Its 
spectral energy distribution \citep{Dewangan&Anandarao10} and its 
 distinctive nature as a submillimeter source 
shows that s11 is a YSO. Nevertheless,
source s11 is clearly offset from the center of the outflow, and it is
hard to imagine that it could be its main driving source. However, we
note that both, the redshifted and blueshifted CO lobes show extensions
to the east \citep{Gomez-etal92}. We suggest that these extensions
could in fact be part of an independent outflow driven by source s11,
while the bulk of the CO outflow would be driven by the sources in the
central cluster.

\subsection{Sources of mass loss in the region}

There are several sources in AFGL 437 that could be undergoing
mass-loss simultaneously. Source WK 34 is associated with an infrared
polarized nebula \citep{Weintraub&Kastner96,Meakin-etal05}, and
has been suggested to be the dominant exciting source of the
outflow. The infrared nebula and its association with a water maser
\citep{Torrelles-etal92,Weintraub&Kastner96} clearly indicate that
this source is an active source of mass-loss.  Source AFGL 437S also
shows elongated infrared emission 
\citep{Alvarez-etal04}, which is also suggestive of a star undergoing
mass-loss, although this extension is weak, and close to the
resolution limit of the maps.

Our  radio continuum map at 2 cm (Fig. \ref{radio3}) shows, for the first
time, the presence of a collimated jet associated to source AFGL
437W. The previous detection of both radio continuum and water maser
emission (Torrelles et al. 1992) already signaled this source as a
young object, but now we see that its mass loss is highly
collimated.

Therefore we have strong evidence that at least WK 34 and AFGL 437W
(and possibly AFGL 437S)
are undergoing collimated mass loss. This seems not to support the
model of \cite{Gomez-etal92}, which suggested that the
mass loss could be isotropic (and the observed CO bipolarity would be
a projection effect). We now favor that the observed CO outflow is the
superposition of several bipolar outflows, which would explain its low
degree of collimation. Sources WK 34, AFGL 437W 
 (and possibly AFGL 437S) could be
responsible for the bulk of high-velocity CO emission close to the
center. We also propose that source s11 is a young embedded object
that could drive an additional outflow, traced by the eastern
extensions of the CO lobes. \cite{Weintraub&Kastner96} and 
\cite{Dewangan&Anandarao10} noticed that the infrared emission near WK 34 is
oriented in the N-S direction next to the source, but it bends to the
northeast away from it. \cite{Dewangan&Anandarao10} suggested that this 
bending may be due to the interaction  of the outflow from WK 34 with 
mass loss from source AFGL 437N, but we suggest that it could
reflect the interactions between the outflows from WK 34 and AFGL
437W.

 This superposition of several individual outflows,
giving rise to complex morphologies when observing at low
angular resolution, is also found in other high-mass star forming regions
(e.g., Beuther et al. 2002, 2006; Ginsburg et al. 2009).

An alternative scenario would be that only one source is significantly
driving the (low-collimation) outflow, even when present evidence is of
highly collimated jets, if both a collimated jet and a low-collimation
wind are driven simultaneously by the same source. Evidence for
simultaneous presence of high- and low-collimation mass loss has been
found for the first time in a high-mass young star in water maser
observations of Cepheus A \citep{Torrelles-etal11}.

To test these scenarios, we propose that
the molecular outflow should be observed with
millimeter/submillimeter interferometers, at scales of a few arcseconds, to be able
to discriminate the possible individual outflows driven by these
sources. Being able to trace the molecular outflow close to its
driving source(s) would provide direct evidence of whether there is a
superposition of outflows or if one source is responsible for it.

\section{Conclusions}
We presented new interferometric NH$_3$(1,1) and (2,2) observations of
the  region of high-mass star formation  AFGL 437, with the aim to
understand the nature of the low-collimation  bipolar molecular
outflow previously detected in CO (G\'omez et al. 1992). These
observations were complemented with archive data of radio continuum
emission at 2 cm, 3.6 cm, and 450 $\mu$m.
Our main conclusions are as follow:

\begin{enumerate}
\item  
 
 The ammonia emission in our interferometric maps is  
located mainly in three clumps at
 the south and east of the central cluster. 

\item The analysis of the  velocity pattern of the ammonia transitions 
in the whole region is not conclusive and,  in principle, it does not allow  us to accept 
or reject the model proposed by G\'omez et al. (1992).

\item One of the clumps seen in NH$_3$(1,1) coincides with the
 maximum of NH$_3$(2,2) emission, and with a local maximum of emission
 at 450 $\mu$m. A near infrared source  (s11) is also found at
 that position. We identify this source as {\em possible} active  young stellar
 object, and it might
 be a source of mass-loss in the AFGL 437 region.

\item The radiocontinuum map at 2 cm shows extended, elongated 
 emission in the source AFGL 437W, not  detected in previous studies.
 This elongated morphology and its spectral index ($\simeq 0.4$) suggest the presence
of a jet in that source.

\item The  collimated mass loss observed in sources WK 34 and AFGL 
437W sources does not support the model proposed by \cite{Gomez-etal92}, 
which suggested an isotropic mass loss in the region.  

\item We suggest that   several bipolar outflows may exist in the AFGL 
437 region. The  CO outflow  would be the superposition of them, what 
would explain its  low degree of collimation. An alternative scenario 
is that the CO outflow could be produced by a main source driving 
both less and highly collimated winds. Observations with millimeter/submillimeter 
interferometers to scales of several arc seconds may identify the individual outflows.

\end{enumerate}

\section*{Acknowledgments}

GM, JFG and IdG acknowledge partial support from Ministerio de Ciencia
e Innovaci\'on (Spain), grant AYA2008-06189-C03-01. JFG is also
supported by Junta de Andaluc\'{\i}a (TIC-126).
This research used the facilities of the Canadian Astronomy Data
Centre operated by the the National Research Council of
Canada with the support of the Canadian Space Agency.

\label{lastpage}

\end{document}